# Cluster glass magnetism in the phase-separated $Nd_{2/3}Ca_{1/3}MnO_3$ perovskite

Elena Fertman*, Sergiy Dolya, Vladimir Desnenko, and Anatoly Beznosov,

B. Verkin Institute for Low Temperature Physics and Engineering NASU, 47 Lenin Ave.,

Kharkov 61103, Ukraine

Marcela Kajňaková and Alexander Feher

P. J. Šafárik University in Košice, Faculty of Science,

Park Angelinum 9, 04154 Košice, Slovakia

## Abstract

A detailed study of the low-temperature magnetic state and the relaxation in the phase-separated colossal magnetoresistance $Nd_{2/3}Ca_{1/3}MnO_3$ perovskite has been carried out. Clear experimental evidence of the cluster-glass magnetic behavior of this compound has been revealed. Well defined maxima in the in-phase linear ac susceptibility $\chi'(T)$ were observed, indicative of the magnetic glass transition at $T_g \sim 60$ K. Strongly divergent zero-field-cooled and field-cooled static magnetizations and frequency dependent ac susceptibility are evident of the glassy-like magnetic state of the compound at low temperatures. The frequency dependence of the cusp temperature $T_{max}$ of the $\chi'(T)$ susceptibility was found to follow the critical slowing down mechanism. The Cole-Cole analysis of the dynamic susceptibility at low temperature has shown extremely broad distribution of relaxation times, indicating that spins are frozen at "macroscopic" time scale. Slow relaxation in the zero-field-cooled magnetization has been experimentally revealed. The obtained results do not agree with a canonical spin-glass state and indicate a cluster glass magnetic state of the compound below $T_g$, associated with its antiferromagnetic-ferromagnetic nano-phase segregated state. It was found that the relaxation mechanisms below the cluster glass freezing temperature $T_g$ and above it are strongly different. Magnetic field up to about $\mu_0 H \sim 0.4$ T suppresses the glassy magnetic state of the compound.

*Corresponding author. Elena L. Fertman, Institute for Low Temperature Physics and Engineering, 47 Lenin Ave., Kharkov 61103, Ukraine. Email address: fertman@ilt.kharkov.ua; Fax: 38-057-3450593.

## 1. Introduction

Insulating $Nd_{2/3}Ca_{1/3}MnO_3$ perovskite is a colossal magnetoresistance narrow-band perovskite compound which possesses phase-segregated state at low temperatures. Earlier we have found that its low temperature magnetic nanoscale phase-segregated state [1, 2] originates from the charge ordering that takes place at much higher temperature $T_{CO} \approx 212$ K [3]. Charge ordering phase transformation in $Nd_{2/3}Ca_{1/3}MnO_3$ is a first order phase transition of martensitic (non-diffusion) type which leads to the self-organized coexistence of charge ordered and charge disordered phases in the wide temperature range below the room temperature. Extended temperature hysteresis of the magnetic susceptibility in the charge ordering region is one of the evidences of the nonequilibrium phase-segregated state [3]. Below $T_{CO}$ the compound exhibits a sequence of magnetic transformations: two antiferromagnetic ones at $T_{N1} \sim 130$ K and $T_{N2} \sim 80$ K, and a ferromagnetic one at $T_C \sim 70$ K. They lead to the coexistence of at least three different magnetic phases at low temperatures: the two antiferromagnetic charge ordered ones (AFM) (their fraction is about 82%) and the ferromagnetic charge disordered (FM) one (its phase fraction is about 18%) [4]. Direct observation of magnetic domains shows that fine ferromagnetic particles are embedded in an antiferromagnetic matrix [5].

In recent years, unusual nonequilibrium dynamics and time-dependent phenomena in phase-segregated perovskites have been reported [6, 7, 8, 9, 4]. In particular, slow relaxation and memory effects in magnetization, the frequency-dependent ac susceptibility and aging were found. The glassy behavior in the phase-separated perovskites is strongly associated with the coexistence of different magnetic phases and therefore, such compounds do not behave like canonical spin glasses, they rather have to be considered as self-organized assembly of interacting magnetic clusters.

In the present work we have investigated magnetic behaviors of the nano-phase segregated $Nd_{2/3}Ca_{1/3}MnO3$ perovskite, and we have shown that below long-range AFM and FM ordering temperatures the compound exhibits one more magnetic transition, namely a cluster-glass freezing one, which occurs at $T_g \sim 60$ K. It does not lead to a typical spin glass state observed in a variety of disordered magnetic systems, but rather to a cluster-glass one associated with the coexistence of AF and FM magnetic phases.

## 2. Experiment

A ceramic $Nd_{2/3}Ca_{1/3}MnO_3$ sample was prepared by a standard solid state reaction technique from stoichiometric amounts of proper powders. X-ray crystal-structure analysis indicated a single-phase material at the room temperature.

Magnetic measurements were made using Quantum Design Magnetic Properties Measurement System (MPMS) and a noncommercial superconducting quantum interference device (SQUID) magnetometer. Zero-field-cooled (ZFC) and field-cooled (FC) dc magnetizations were measured at various applied magnetic fields up to 5 T in the temperature range 2 – 300 K. The in-phase, $\chi'(T)$, and out-of-phase, $\chi''(T)$, components of the complex ac susceptibility $\chi(T) = \chi'(T) - i\chi''(T)$, were measured at different frequencies from 1.3 Hz to 1116 Hz at fixed ac field 0.25 mT in warming runs by MPMS. The temperature was stabilized within 0.1 K at each measuring point.

## 3. Results and discussion

### 3.1. Static magnetization

Temperature dependences of dc magnetization for $Nd_{2/3}Ca_{1/3}MnO_3$ in zero-field-cooled $M_{ZFC}(T)$ and field-cooled $M_{FC}(T)$ conditions taken in magnetic fields $\mu_0H$ from 0.002 to 5 T are shown in Fig. 1. The FC and ZFC magnetizations have the same value at high temperatures and they begin to diverge strongly at low temperatures. Such difference between the FC and ZFC magnetization is typical for both classical spin glasses (SG) [10, 11] and phase-separated manganite systems with coexisting FM clusters and an AFM matrix [12, 13]. A sharp maximum $T_{max}$ in $M_{ZFC}$ seen at low magnetic fields (Fig. 1 c) is also reminiscent of spin-glass or cluster-glass systems as well (CG) [14]. This maximum broadens and shifts to the low temperatures with applied magnetic field (Fig. 1 a, b).

Anomalies of the temperature dependent ZFC and FC magnetization curves (Fig. 1) as well as anomalies of the dynamic susceptibility temperature dependent curves (Fig. 2) below 30 K are driven by the ordering of the $Nd^{3+}$ ions magnetic subsystem [4].

The ZFC magnetization cusp of about 60 K, measured at the lowest magnetic field ($\mu_0H = 0.002$ T) may be considered as the spin-glass freezing temperature $T_g$. The form of the $M_{ZFC}(T)$ curve measured in low magnetic field and its cusp coincide well with the cusp seen in the initial ac susceptibility $\chi'(T)$ at the lowest measured frequency (1.3 Hz).

Canonical SG systems typically display a bifurcation of the ZFC and FC magnetizations very close to the SG freezing temperature $T_g$ [10], but in our case the point at which bifurcation occurs, the irreversibility temperature $T_{irr}$, is much higher than $T_{max}$: in magnetic field $\mu_0 H = 0.002$ T we found $T_{irr} \sim 85$ K, $T_{max} \sim 60$ K (Fig. 1). Reason of this may be cluster-glass magnetic behavior [15, 7, 12].

Taking into account the phase-segregated state of the compound, the temperature dependent dc magnetic susceptibility may be interpreted as follows. Upon zero-field cooling the ferromagnetic clusters, which exist below 70 K in the phase-separated $Nd_{2/3}Ca_{1/3}MnO_3$ compound [1, 2, 4], freeze into random orientations, as dictated by the local anisotropy field, while field-cooled clusters align, leading to the onset of a magnetization.

Applied magnetic field substantially changes the character of the $M_{ZFC}(T)$ curve (Fig. 1), shifting the cusp temperature of ZFC magnetization $T_{max}$ and diverging temperature $T_{irr}$ closer to each other (Fig. 2). For low magnetic fields $\mu_0 H < 0.3$ T we found that $\Delta T = T_{irr} - T_{max}$ changes relatively slowly; this suggests that the applied field does not change the size of the ferromagnetic clusters substantially, and the increasing magnetization comes from an alignment of their moments with the increasing field. For low-field region the effect of the applied magnetic field on the magnetic state of the studied compound was estimated by fitting the field-dependent $T_{irr}(H)$ and $T_{max}(H)$ by the expressions:

$$T_{irr}(H) = T_{irr}(0) \cdot \left(1 - \left(H/H_{irr}\right)^p\right) \tag{1}$$

$$T_{max}(H) = T_{max}(0) \cdot \left(1 - \left(H/H_g\right)^q\right) , \tag{2}$$

where $T_{irr}(0) = 100$ K, $H_{irr} = 0.3$ T, $p = 0.5$; $T_{max}(0) = 61$ K, $H_g = 0.4$ T, $q = 0.5$.

The obtained parameters $H_{irr}$ and $H_g$ are very similar, and they were treated as "glassy" parameters, pointing to the magnetic fields which effectively suppress the glassy magnetic state of the compound. Their values well agree with the threshold field of a field-induced AF-FM transition $H_f \sim 0.4$ T found in $Nd_{2/3}Ca_{1/3}MnO_3$ at low temperatures [16].

The obtained value of the parameter $p = 0.5$ is evident more of the cluster-glass behavior, than of the canonical spin-glass one, for which p = 2/3 is characteristic; p < 2/3 is characteristic for the cluster glass formation [17, 18].

For high-field regime $\mu_0 H > 0.4$ T, $\Delta T$ is approximately zero. At the same time the form of the ZFC magnetization curve changes drastically, indicating an increase of the ferromagnetic fraction due to the growth of the ferromagnetic cluster.

Thus the obtained dc data appear to support the idea of the cluster-glass ground state of the compound rather that of the conventional spin-glass state: a) the irreversibility temperature $T_{irr}$ and the glass-freezing temperature $T_{max}$ do not coincide in the applied low magnetic fields; b) the approximations of $T_{irr}(H)$ and $T_{max}(H)$ in low magnetic fields $\mu_0 H < 0.3$ T give the parameter value $p = 0.5$ which agrees with the cluster-glass behavior.

*3.2. Dynamic magnetization*

In Fig. 3 the temperature dependences of the real ($\chi'$) and imaginary ($\chi''$) components of the ac susceptibility of $Nd_{2/3}Ca_{1/3}MnO_3$ at frequency range of 1.3 to 1116 Hz are shown. The $\chi'(T)$ curves display a rather sharp peak at temperature $T_{max}$ which shifts upwards with increasing frequency (Fig. 3a, the inset). This is a distinct feature of both the spin-glass and the cluster-glass magnetic states [10, 11, 14]. Note that $\chi''(T)$ curves demonstrate one more peak below about 30 K; the $\chi'(T)$ curves have the corresponding bend anomaly. This low temperature anomaly is attributed to the magnetic ordering of $Nd^{3+}$ moments (neodymium subsystem) and has been discussed in paper [4].

We have analyzed the frequency dependence of the $T_{max}$ in terms of the dynamical slowing down of the spin fluctuations above the glass transition temperature in a three-dimensional spin glass [19] as described by

$$\tau = 1/(2\pi f) = \tau_0 (T_{\max}/T_g - 1)^{z\nu} ,  \qquad (3)$$

where $T_g$ is the critical temperature for spin glass ordering (this is equivalent to the $f \to 0$ of $T_{max}$), $z\nu$ is a constant exponent, $\tau_0$ is the characteristic time scale for spin dynamics (the shortest relaxation time of the system).

The least squares fitting of the frequency dependence of $T_{max}$ (Fig. 4) corresponding to the peak temperature in the $\chi'(T)$ curves by Eq. 3 has been done. The best fitting shown in Fig. 4 gives the following values of parameters: $T_g = 58.7$ K, $z\nu = 6.53$, and $\tau_0 = 2.2 \cdot 10^{-14}$ s. Logarithmic plot of $\tau / \tau_0$ vs $(T_{\max} / T_g - 1)$ for $Nd_{2/3}Ca_{1/3}MnO_3$ (Fig. 4, the inset) demonstrates a good agreement with Eq. 3 and confirms the correctness of the chosen model.

The obtained $T_g$ value is clearly reasonable: it is close to the experimental $T_g = 60$ K obtained from our static magnetic measurements. Other fitting parameters are within the range of typical values for known spin glasses as well as the cluster-glass phase-separated systems, for which $z\nu \sim 6 - 12$, and $\tau_0 \sim 10^{-12} - 10^{-15}$ s [14, 20, 21, 22, 23].

However, the peak in $\chi''(T)$ increases with decreasing frequency. This is qualitatively different from the behavior of most spin glasses in which we expect an increase of the peak magnitude with increasing frequency [12].

The obtained rate of the frequency shift $\dfrac{dT_{\max}}{T_{\max} \cdot d\log_{10}(f)} = 0.0038$ is little lower than a typical value for canonical spin systems in which it ranges from 0.0045 to 0.28 [10] and it is rather consistent with a cluster-glass behavior. However, this rate is somewhat higher than the value for the related $Pr_{2/3}Ca_{1/3}MnO_3$ compound where it 0.00154 [12].

Thus, the obtained ac magnetization data, namely the value of the obtained frequency shift, and the decrease of the peak magnitude of $\chi''(T)$ with increasing frequency support more the idea of the cluster-glass system than that of classic spin-glass one.

### 3.3. Relaxation processes

To investigate the relaxation behavior of the compound, we performed the Cole-Cole plot analysis for spin glass systems [24, 25]. The complex susceptibility can be phenomenologically expressed as

$$\chi = \chi_s + \frac{\chi_0 - \chi_s}{1 + \left(i\omega\tau_C\right)^{1-\alpha}} \quad , \tag{4}$$

where $\chi_0$ and $\chi_s$ are the isothermal ($\omega = 0$) and adiabatic ($\omega \to \infty$) susceptibilities, respectively, $\tau_C$ is the median relaxation time around which the distribution of relaxation times is assumed, while $\alpha$ ($0 < \alpha < 1$) is representative of the width of the distribution. From the analysis the distribution function of relaxation times at each temperature can be determined. The Argand diagrams for the studied compound are shown in Fig. 5, where the $\chi'$ and $\chi''$ from Fig. 3 are plotted in the complex plane. These diagrams appear to be very unusual: there are no maxima on the diagrams at temperatures below freezing temperature $T_g \sim 60$ K. The "flatness" of the curves is a measure of the distribution width of the relaxation times. So, the obtained data provides an evidence of a very broad distribution of relaxation times while the median relaxation time is shifted to very large values at all studied temperatures, indicating that spins are frozen at "macroscopic" time scale. This strongly differs from canonical spin glasses [24, 25] and resembles a cluster-glass behavior [25]. We observe that at temperatures above $T_g$, the distribution of relaxation times becomes less broad indicating the "usual" spin dynamics, while the maxima of the diagrams appear to indicate the median relaxation time (Fig. 5, the inset). It implies that magnetic moments of clusters are frozen at "macroscopic" time scales below $T_g$ and they fluctuate above this temperature.

To further investigate the glassy magnetic behavior, we have measured the relaxation in ZFC magnetization. The sample was cooled from 300 to 10 K at 4 K/min in zero magnetic field. Then the sample was warmed to a measuring temperature and kept at this temperature for a half an hour to allow the phase-separated state to relax. Afterwards the magnetic field $\mu_0H = 10$ mT was applied and the magnetization data were collected immediately after the magnetic field value was set.

Figure 6 shows the ZFC magnetization *M(t)* and the normalized magnetization *M(t)/M*(0) as a function of time. The magnetization data were collected at 10 K and 79 K. At 10 K there is a continuous increase of the magnetization even after 2.5 hours, during which about 25% of the FC magnetization (in $\mu_0 H = 10$ mT) were reached. We have fitted the low-temperature *M*(t) dependence as an exponential decay of the magnetization [26]:

$$M(t) = M(\infty) \cdot \left[ 1 - \beta \cdot \exp\left( -\left( (t / \tau)^{\alpha} \right) \right) \right],$$ (5)

where $M(\infty) = \lim\limits_{t \to \infty} M(t)$ and $\tau$ is the characteristic relaxation time dependent on the temperature *T*. We have obtained following fitting parameters: *M*(∞) = 0.38 emu/g, $\beta$ = 0.184, $\tau$ = 24.9 min, $\alpha$ = 0.287. It needs to be noted that the obtained characteristic relaxation time $\tau$ = 24.9 min agrees well with the one obtained for the field-cooled magnetization relaxation $\tau$ = 29 min [27].

The long-time relaxation behavior observed at low temperatures is associated with the tendency of ferromagnetic clusters to align in the applied magnetic field. This agrees well with the very broad distribution of relaxation times resulting from the Cole-Cole analysis of dynamic magnetization, indicating that magnetic moments of coexisting clusters are frozen at macroscopic time scale. These results strongly disagree with the canonical spin-glass behavior which is what we have assumed [9].

The time dependent magnetization *M(t)* measured at temperature of 79 K, which is above the freezing temperature $T_g$, is very unusual. The *M(t)* shows a positive slope at the beginning (∼ 20 min) of the time run, indicating that the FM contribution associated with the alignment of the moments of ferromagnetic clusters rises. At this temperature, the magnetization increases with time much less than at 10 K, this is associated with a decreased ferromagnetic fraction. The magnetization reaches its maximum after approximately 20 minutes and then starts to decrease (Fig. 6). This curious peak of *M(t)* observed is related to the competition between two different low magnetic field relaxation processes, the "up" relaxation one and the "down" relaxation one. The "up" relaxation process dominates in the beginning; the "down" one dominates after about 20 min at this temperature. The "up" process, was fitted by the equation 5, the same way as was done at 10 K. The best fitting parameters are *M*(∞) = 0.57 emu/g, $\beta$ = 3.055, $\tau$ = 174 min, $\alpha$ = 0.114. The nature of the "down" relaxation process is not quite clear so far and needs to be further studied. It is presumably caused by the processes occurring on the interfaces between ferromagnetic clusters and antiferromagnetic matrix, their thermally activated dynamics and non equilibrium magnetic state near the Curie temperature ($T_C \sim 70$ K). We think that the unusual magnetic relaxation found at 79

K can not be attributed to external factors, caused by relaxation in the magnet or in nearby magnetic materials in the building [28], as our relaxation measurements have been done by home made SQUID magnetometer, where the magnetic field $\mu_0 H = 10$ mT was frozen and controlled by a Hall sensor. So, we have no uncontrolled residual field as opposed to [28].

The long-time relaxation behavior found at low temperatures, as well as our measurements of static and dynamic magnetizations, are not consistent with a typical spin-glass phase observed in a variety of disordered magnetic systems. They are consiste with a cluster magnetic glass state of the compound [9] and strikingly remind the magnetic behavior and an unusual long relaxation time of interacting magnetic nanoparticle systems [29, 30].

## 4. Conclusions

Low temperature magnetization study of the colossal magnetoresistance $Nd_{2/3}Ca_{1/3}MnO_3$ compound has been done. Clear experimental evidences of the cluster-glass magnetic behavior of the compound below the freezing temperature $T_g \sim 60$ K have been revealed. Slow relaxation process and extremely broad distribution of relaxation times found, indicating that spins are frozen at "macroscopic" time scale, confirm the cluster-glass magnetic state, associated with the antiferromagnetic-ferromagnetic phase segregated state of the compound. The relaxation processes above the freezing temperature $T_g$ and below it have been found strongly different. Magnetic field up to about $\mu_0 H \sim 0.4$ T suppresses the glassy magnetic state of the compound.

## 5. Acknowledgements

Authors are thankful to Dr. D. Sheptyakov (PSI, Switzerland) and to Dr. D. Khalyavin (ISIS, United Kingdom) for fruitful collaboration.

This work was supported by: SAS Centre of Excellence CFNT MVEP, by the ERDF EU (European Union European regional development fond) grant under the contract No.TMS26220120005 and by Slovak Grant Agency VEGA-1/0159/09. The financial support of U. S. Steel Košice is gratefully acknowledged.

**Figure captions**

Fig. 1. ZFC (full symbols) and FC (open symbols) static magnetizations for $Nd_{2/3}Ca_{1/3}MnO_3$ as a function of temperature, measured in the magnetic field ranges $\mu_0 H = 0.002 - 5$ T.

Fig. 2. The field dependence of the analyzed static magnetization data of the $Nd_{2/3}Ca_{1/3}MnO_3$ compound, the irreversibility temperature $T_{irr}$ and the cusp of ZFC magnetization $T_{max}$; solid lines are their approximations $T_{irr}(H) = 100 \cdot \left(1 - \left(H/0.3\right)^{0.5}\right)$ and $T_{max}(H) = 61 \cdot \left(1 - \left(H/0.4\right)^{0.5}\right)$, correspondingly.

Fig. 3. Real ($\chi'$) and imaginary ($\chi''$) components of the ac susceptibility of $Nd_{2/3}Ca_{1/3}MnO_3$ measured at different frequencies.

Fig. 4. Frequency dependence of $T_{max}$ of the $Nd_{2/3}Ca_{1/3}MnO3$ compound, corresponding to the peak temperature in the $\chi'(T)$ curves (Fig. 3) and the corresponding fit by Eq. 3. Inset: Logarithmic plots of $\tau/\tau_0$ vs $(T_{max}/T_g - 1)$ for $Nd_{2/3}Ca_{1/3}MnO_3$, demonstrating the agreement with Eq. 3.

Fig. 5. Argand diagrams of $Nd_{2/3}Ca_{1/3}MnO_3$ for different temperatures below freezing temperature $T_g \sim 60$ K and above this temperature (the inset). The lines are fits by Eq. 4.

Fig. 6. The time dependence of ZFC magnetization (a) and normalized magnetization $M(t)/M(0)$ (b) for $Nd_{2/3}Ca_{1/3}MnO_3$ at 10 K and 79 K. The magnetization data were collected immediately after applying the magnetic field $\mu_0 H = 10$ mT. Solid lines (red in color) are the best fits of the magnetic relaxation by Eq. 5. Dash lines are guide for eyes only. The peak on the 79 K curves is related to the competition between two different relaxation mechanisms.

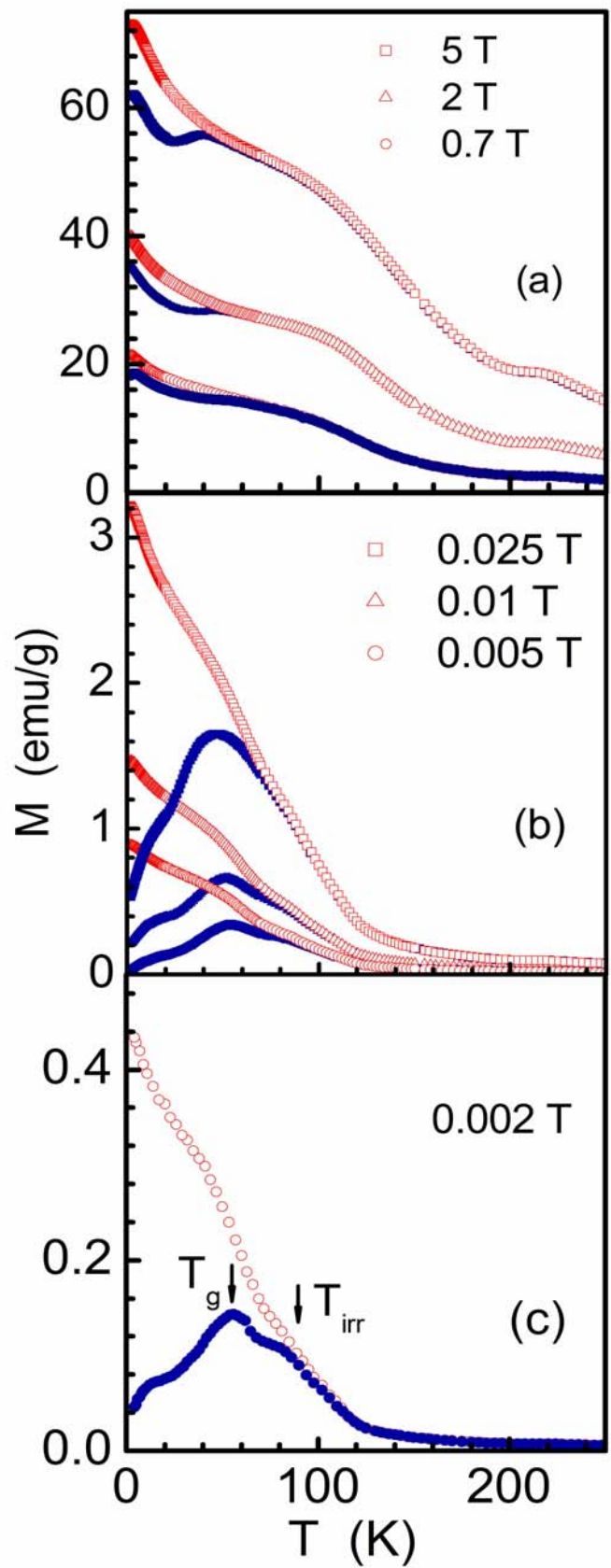

Fig. 1. ZFC (full symbols) and FC (open symbols) static magnetizations for $Nd_{2/3}Ca_{1/3}MnO_3$ as a function of temperature, measured in the magnetic field ranges $\mu_0 H = 0.002 - 5$ T.

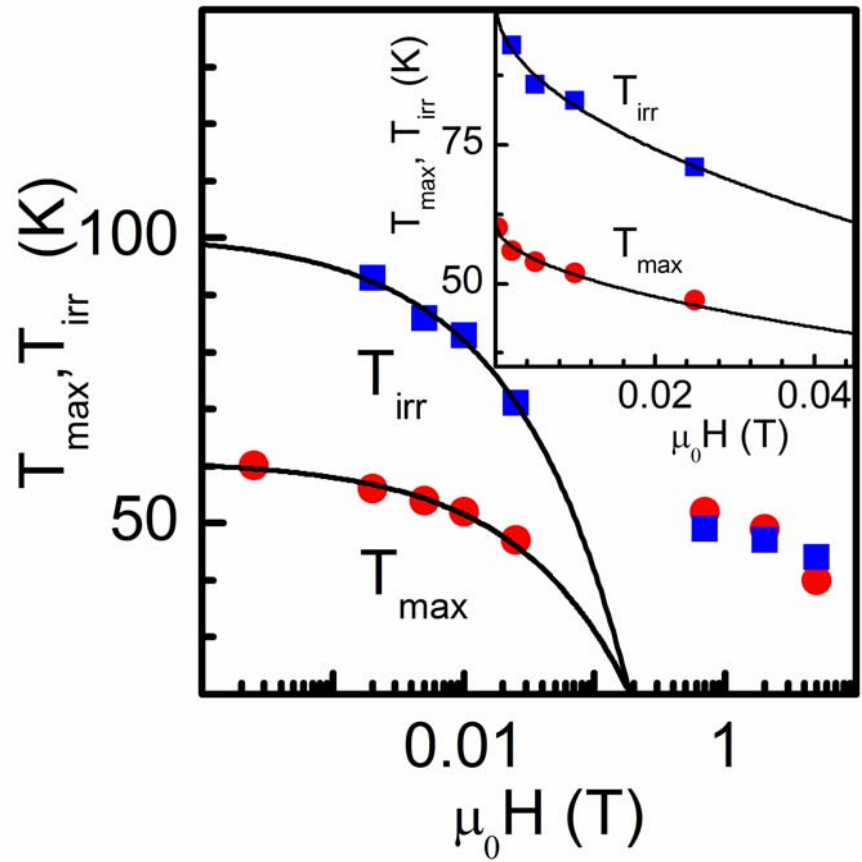

Fig. 2. The field dependence of the analyzed static magnetization data of the $Nd_{2/3}Ca_{1/3}MnO_3$ compound, the irreversibility temperature $T_{irr}$ and the cusp of ZFC magnetization $T_{max}$; solid lines are their approximations $T_{irr}(H) = 100 \cdot \left(1 - \left(H/0.3\right)^{0.5}\right)$ and $T_{max}(H) = 61 \cdot \left(1 - \left(H/0.4\right)^{0.5}\right)$, correspondingly.

E. Fertman, S. Dolya, V. Desnenko, A. Beznosov, M. Kajnakova, and A. Feher, Cluster-glass magnetism in the phase-separated $Nd_{2/3}Ca_{1/3}MnO_3$ perovskite

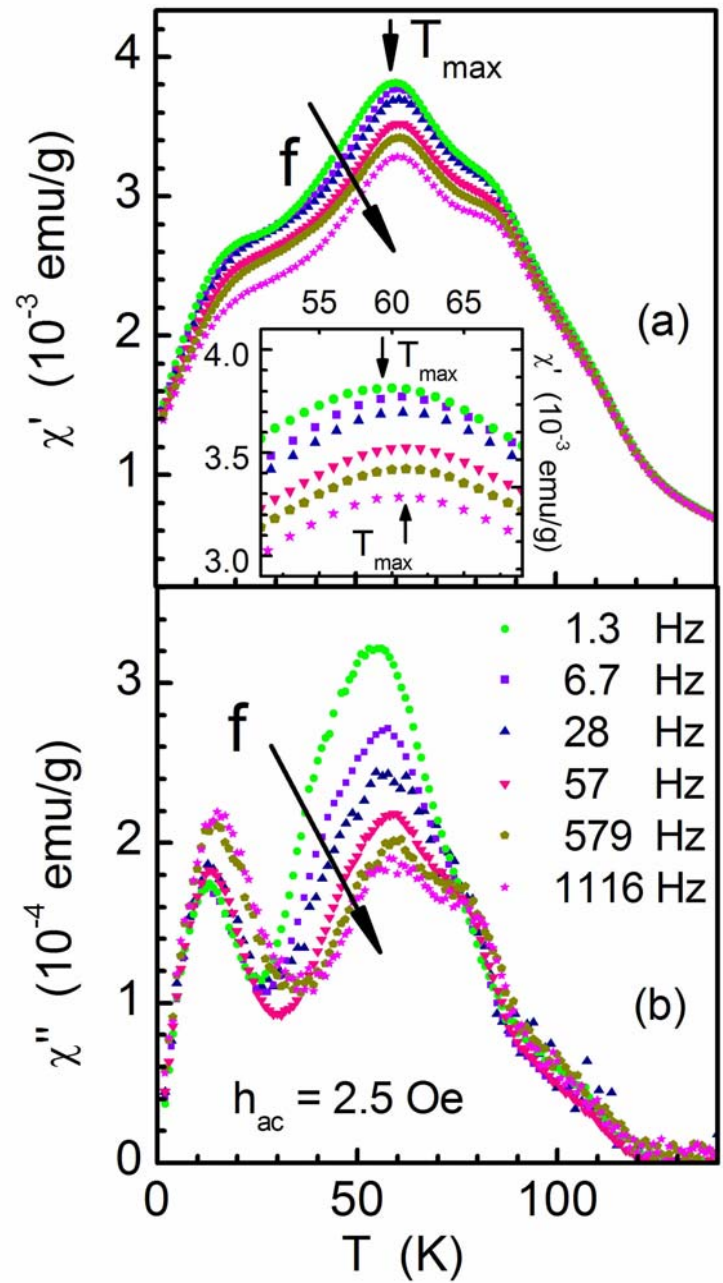

Fig. 3. Real ($\chi'$) and imaginary ($\chi''$) components of the ac susceptibility of Nd$_{2/3}$Ca$_{1/3}$MnO$_3$ measured at different frequencies.

E. Fertman, S. Dolya, V. Desnenko, A. Beznosov, M. Kajnakova, and A. Feher, Cluster-glass magnetism in the phase-separated Nd$_{2/3}$Ca$_{1/3}$MnO$_3$ perovskite

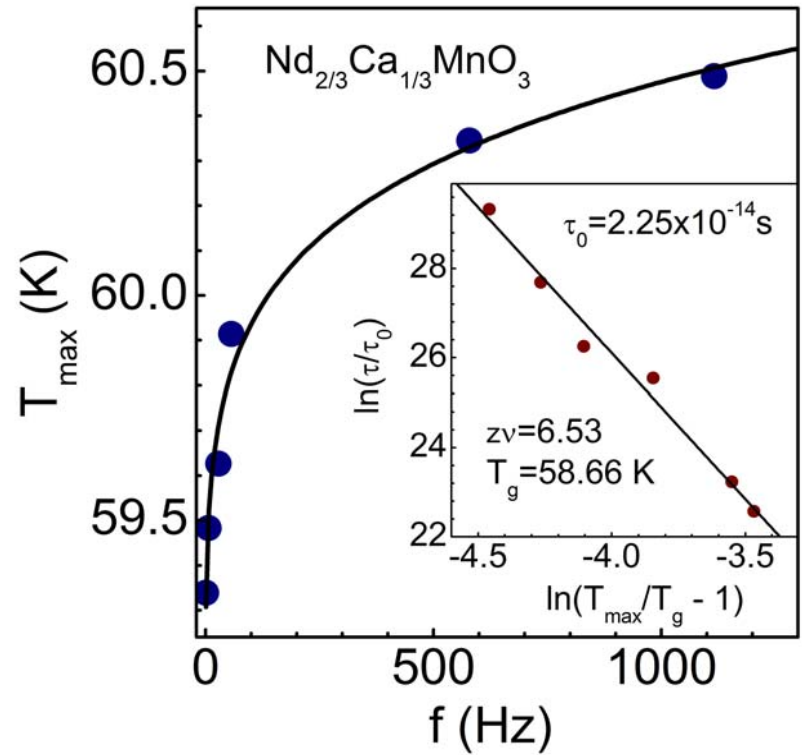

Fig. 4. Frequency dependence of $T_{max}$ of the $Nd_{2/3}Ca_{1/3}MnO_3$ compound, corresponding to the peak temperature in the $\chi'(T)$ curves (Fig. 3) and the corresponding fit by Eq. 3. Inset: Logarithmic plots of $\tau/\tau_0$ vs ($T_{max}/T_g$ - 1) for $Nd_{2/3}Ca_{1/3}MnO_3$, demonstrating the agreement with Eq. 3.

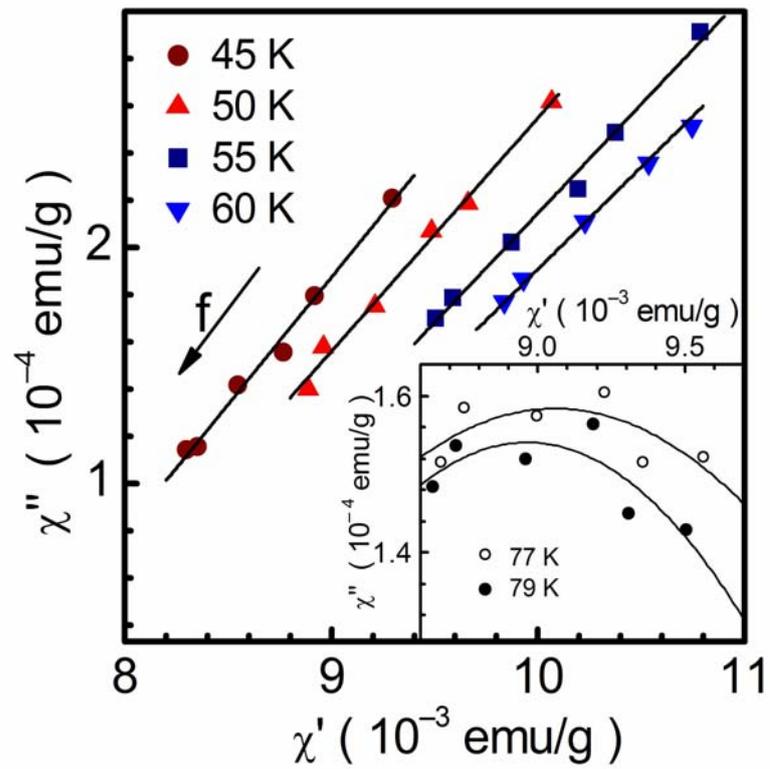

Fig. 5. Argand diagrams of Nd$_{2/3}$Ca$_{1/3}$MnO$_3$ for different temperatures below freezing temperature $T_g \sim 60$ K and above this temperature (the inset). The lines are fits by Eq. (4).

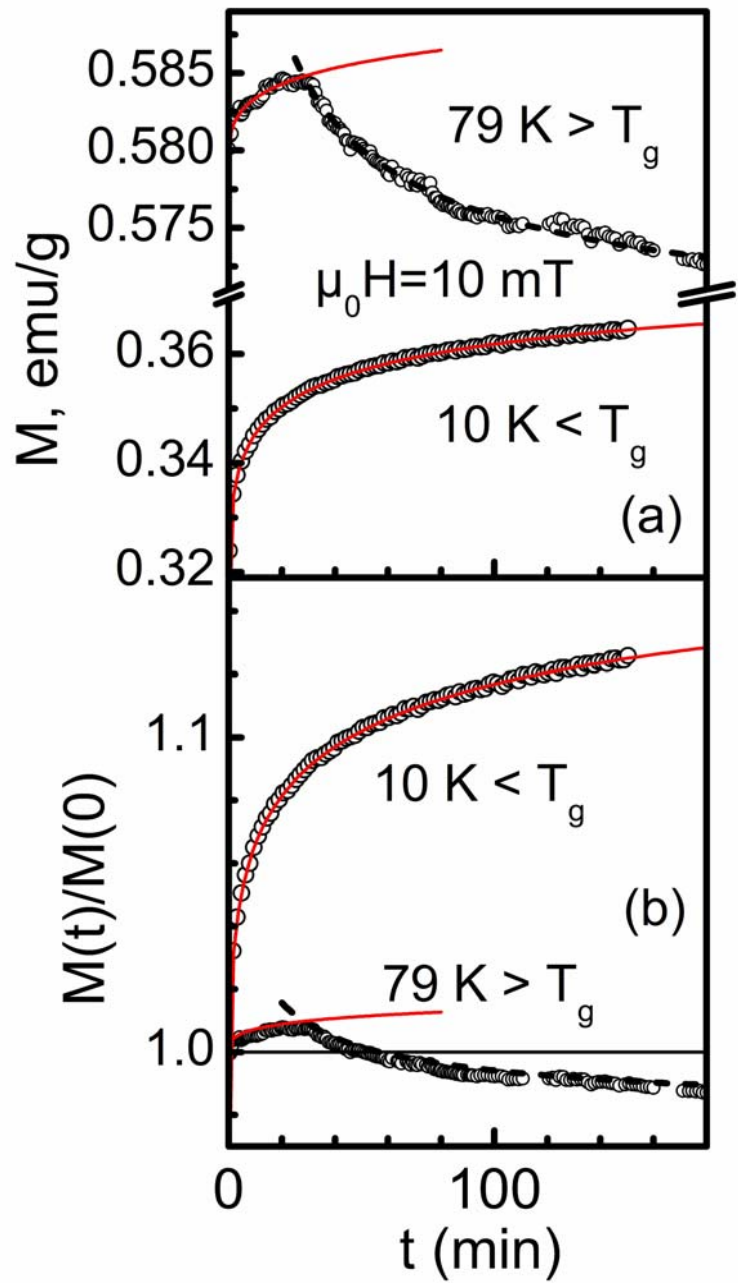

Fig. 6. The time dependence of ZFC magnetization (a) and normalized magnetization *M(t)/M(0)* (b) for Nd$_{2/3}$Ca$_{1/3}$MnO$_3$ at 10 K and 79 K. The magnetization data were collected immediately after applying the magnetic field *μ$_0$H* = 10 mT. Solid lines (red in color) are the best fits of the magnetic relaxation by Eq. 5. Dash lines are guide for eyes only. The peak on the 79 K curves is related to the competition between two different relaxation mechanisms.